\documentclass[aps,twocolumn,superscriptaddress,showpacs]{revtex4}

\usepackage{graphicx}% Include figure files
\usepackage{dcolumn}% Align table columns on decimal point
\usepackage{bm}% bold math
\usepackage{amsmath}
\usepackage{epsfig}

\begin{document}

%\preprint{APS/123-QED}

\title{Competition between Intra-community and Inter-community Synchronization}

\author{Ming Zhao$^{1,2,3}$}
\email{zhaom17@gmail.com}
\author{Changsong Zhou$^{4}$}
\email{cszhou@hkbu.edu.hk}
\author{Jinhu L\"{u}$^{5}$}
\email{jhlu@iss.ac.cn}
\author{Choy Heng Lai$^{1,2}$}
\email{phylaich@nus.edu.sg}

\affiliation{$^{1}$Department of Physics, National University of Singapore, Singapore 117542 \\
$^{2}$Beijing-Hong Kong-Singapore Joint Center of Nonlinear and Complex Systems (Singapore), Singapore 117542 \\
$^{3}$College of Physics and Technology, Guangxi
Normal University, Guilin 541004, P. R. China\\
$^{4}$Department of Physics, Centre for Nonlinear Studies, and The
Beijing-Hong Kong-Singapore Joint Centre for Nonlinear and Complex
Systems (Hong Kong), Hong Kong Baptist University, Kowloon Tong,
Hong Kong, P. R. China\\
$^{5}$Key Laboratory of Systems and Control, Institute of Systems
Science, Academy of Mathematics and Systems Science, Chinese
Academy of Science, Beijing 100080, P. R. China
}

\date{\today}

\begin{abstract}
In this paper the effects of external links on the synchronization
performance of community networks, especially on the competition
between individual community and the whole network, are studied in
detail. The study is organized from two aspects: the number or
portion of external links and the connecting strategy of external
links between different communities. It is found that increasing
the number of external links will enhance the global
synchronizability but degrade the synchronization performance of
individual community before some critical point. After that the
individual community will synchronize better and better as part of
the whole network because the community structure is not so
prominent. Among various connection strategies, connecting nodes
belonging to different communities randomly rather than connecting
nodes with larger degrees is the most efficient way to enhance
global synchronization of the network. However, a preferential
connection scheme linking most of the hubs from the communities
will allow rather efficient global synchronization while
maintaining strong dynamical clustering of the communities.
Interestingly, the observations are found to be relevant in a
realistic network of cat cortex. The synchronization state is just
at the critical point, which shows a reasonable combination of
segregated function in individual communities and coordination
among them. Our work sheds light on principles underlying the
emergence of modular architectures in real network systems and
provides guidance for the manipulation of synchronization in
community networks.
\end{abstract}

\maketitle

\section{Introduction}

Communities exist ubiquitously in all kinds of networks
\cite{NewmanMEJ2001,Girvan2002,Palla2005} and synchronization on
community networks is of great importance in the biological and
social networks. It has been shown that the sparsity of
connections between different communities hinders the global
synchronization of complex networks
\cite{LHuang2006,Tzhou2007,WangXH2010}. Factors that affect the
synchronization of community networks have been intensively
studied and strategies that can achieve global synchronization
have been proposed
\cite{ParkK2006,GuanSG2008,WangHJ2008,WangKH2009}. Synchronization
transition process of community networks have also been studied
\cite{OhE2005,YanG2007}. Utilizing synchronization to detect
community structure has attracted a great deal of attention
recently
\cite{ArenasA2006,BoccalettiS2007,OhE2008,GfellerD2008,LiXJ2010}.
Moreover, the dynamical modules of networks with and without clear
communities have also been studied in detail
\cite{ArenasA2007,Gomez-GardenesJ2007,BredeM2008,Fuchs2009,ZhaoM2010}.
Very recently, a class of small-world networks with spatial and
network modularity were obtained by evolving the arrangement of
nodes in space and their corresponding network topology
\cite{BredeM2010}. Besides, some other interesting topics, such as
the synchronization interfaces and overlapping communities in
complex networks, were also studied \cite{LiD2008}.

In the study of synchronization on community networks, how the
competition and coordinations of the links within community and
between communities shape the synchronization properties are still
not clearly understood. In this paper, we focus on the effects of
external links (links connecting nodes belong to different
communities) on the synchronization performance of community
networks, especially on the individual community. We study (i) how
the synchronization performances of global network and individual
community are affected by the changing of the number of external
links, (ii) whether connecting nodes with larger degrees of
different communities will ensure better global network
synchronizability, and (iii) the synchronization property of the
cortical network of cat in terms of the above two aspects. This
work will shed light on the understanding of the impacts of the
number and the connecting strategy of external links on the global
network synchronizability and individual community synchronization
performance. This work is organized as follows: in the next
section, we introduce the dynamical equation of each node and the
community network model. We will present the two aspects of
studies on the community network models in the sections III and
IV, and the study on cat cortex in section V. Then we will give
our discussion and conclusion in the last section.

\section{Dynamical Equation and Community Network Model}

To investigate the synchronization on complex networks, dynamical
systems are often taken as nodes and the couplings between
different systems are the links, thus the dynamical equation of
each node in a complex network with $N$ nodes is
\begin{equation}
\dot{\textbf{x}}_i=\textbf{F}(\textbf{x}_i)-\frac{\sigma}{\langle
k \rangle}\sum_{j=1}^N
G_{ij}\textbf{H}(\textbf{x}_j),\hspace*{1em}i=1,...,N,
\end{equation}
where $\dot{\textbf{x}}=\textbf{F}(\textbf{x})$ is the individual
dynamics, $\sigma$ is the overall coupling strength, $\langle k
\rangle$ is the average degree of the network,
$\textbf{H}(\textbf{x})$ is the output function and $G_{ij}$ is
the element of coupling matrix. In our simulations, if node $i$ is
coupled by node $j$, $G_{ij}<0$, otherwise $G_{ij}=0$, and the
diagonal element $G_{ii}=-\sum_{j=1,j\neq i}^N G_{ij}$ to ensure
the sum of the elements in a row is 0 so as to make the
synchronization manifold an invariant manifold. In many previous
work, the global network synchronizability were measured by the
eigenvalues of the coupling matrix without calculating the
iteration of oscillators on the networks according to the master
stability function \cite{Pecora98,Barahona02}. In this paper, we
study not only the global network synchronizability but also the
individual community synchronization performance, thus the master
stability function is insufficient to our study, so we take
R\"{o}ssler oscillator as dynamical node to fulfill our
simulation:
\begin{eqnarray}
\dot{x} &=& -y-z, \nonumber\\
\dot{y} &=& x+ay, \\
\dot{z} &=& b+z(x-c), \nonumber
\end{eqnarray}
where $a=0.2$, $b=0.2$ and $c=7.0$. We take the output function
$\textbf{H}(\textbf{x})=(x,0,0)$. In section II and III, all the
links are un-weighted and the coupling matrix is Laplacian matrix,
i.e., $G_{ij}=-1$ (node $i$ and $j$ are connected) or $G_{ij}=0$
(node $i$ and $j$ are not connected) for off-diagonal elements,
and $G_{ii}=k_i$ for diagonal elements. The correlation between
each pair of nodes $c_{ij}$ is calculated to measure the
synchronization performance, and the average correlation of all
pairs of nodes in the whole network $C_N$ and that of pairs in a
community $C_C$ are taken as the measure of the synchronization
performance of global network and individual community,
respectively.

To implement our study effectively, we design the following
community network model: take $m$ networks as subnetworks to
generate a community network and these subnetworks will be the
communities, then rewire the internal links (links connecting
nodes within the same community) to external links with some
strategies, which will be discussed in detail. When more and more
external links emerge, the modularity will decrease gradually till
the network becomes a homogeneous one.

We follow Ref. \cite{NewmanMEJ2004,NewmanMEJ2004_2} to quantify
modularity of a network with $m$ communities as
\begin{equation}
Q=\sum_l^L (e_{ll}-a_l^2),
\end{equation}
where $a_l=\sum_k^L e_{lk}$ and $e_{lk}$ is the fraction of total
strength of the links in the network that connect the nodes
between the communities $l$ and $k$, namely $e_{lk}=(1/W)
\sum_{i\in l, j \in k} w_{ij}$, with $w_{ij}$ being the connection
strength between two nodes and $W$ the total of $w_{ij}$ in the
network. The topological modularity $Q_T$ measure the strength of
community structure of community networks.
% and is evaluated by taking $w_{ij}=A_{ij}$, the topological matrix of the network.
Synchronization clustering can be quantified by the dynamical
modularity $Q_D$ computed by taking $w_{ij}$ as the dynamical
similarity between two nodes. In particular,
$w_{ij}=(1+c_{ij})/2.0$, where $c_{ij}$ is the correlation between
pair of nodes $i$ and $j$.

\begin{figure}
\scalebox{0.43}[0.43]{\includegraphics{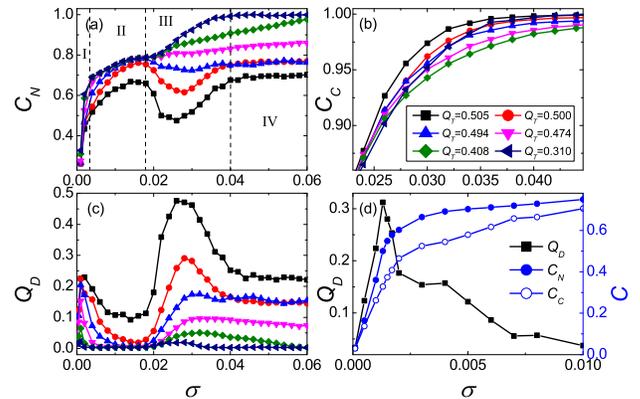}}
\caption{(Color on line) The synchronization performance of global
network $C_N$ (a), of individual community $C_C$ (b), and
dynamical modularity $Q_D$ (c) as functions of the coupling
strength $\sigma$ at different topological modularity $Q_T$ (or
external link number $L_E$). (d) $C_N$, $C_C$ and $Q_D$ at very
small coupling strength for $Q_T=0.500$. There are 2 communities
in the network, each one is a BA scale-free network with 100
nodes, and the average degree in each community is 16. Each plot
is obtained after the averaging over 50 network configurations and
10 initial states of each configuration.}\label{Fig1}
\end{figure}

\section{Effects of external link number}
To study the effects of the number of external links $L_E$ on the
synchronization of global network and individual community, we
take several BA scale-free networks as subnetworks, and construct
community networks with the following operations: (i) random
select an internal link, (ii) cut one end of it and (iii) rewire
it to a random selected node in the other subnetworks, then repeat
the operations (i) to (iii) till the desired community network is
obtained. This connecting strategy is called \emph{random
connection}. We have investigated networks composed by different
number of communities, and found no essential differences on the
synchronization property. Thus, in our simulation we take the
simplest community network model with two communities.

\begin{figure}
\scalebox{0.43}[0.43]{\includegraphics{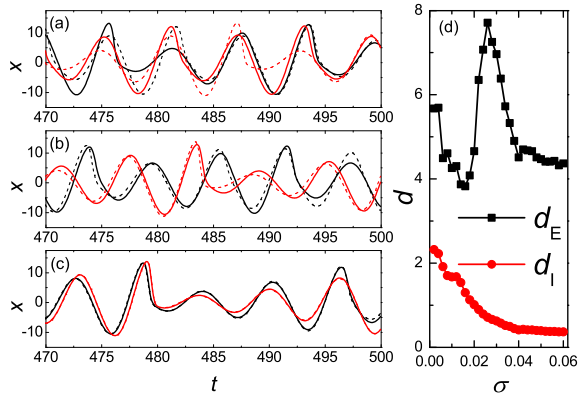}}
\caption{(Color on line) The change of state $x$ with time $t$ at
coupling strength (a) 0.014, (b) 0.026 and (c) 0.060. Two black
curves and two red curves present nodes belong to the two
different communities separately and the two solid curves present
nodes connected by external link and the two dash curves present
two nodes which are neither connected to each other nor to the
other two nodes. (d) The change of $d_E$ (square line) and $d_I$
(circle line) with the coupling strength. The topological
modularity of the adopted network is 0.500 and the other network
parameters are the same as in Fig. \ref{Fig1}.}\label{Fig_1b}
\end{figure}

In Figs. \ref{Fig1}(a) and (b), we presented the changes of $C_N$
and $C_C$ with coupling strength $\sigma$ at different topological
modularity $Q_T$ (corresponding to different number of external
links). Clearly, at the same coupling strength, with the
decreasing of topological modularity, the global network
synchronizability will be better and better, which is consistent
with previous studies \cite{LHuang2006,Tzhou2007,WangXH2010}.
However, the change of global synchronization performance respects
to the coupling strength is of great difference: all the curves
first increase sharply at very small coupling strength [region I
in Fig. \ref{Fig1}(a)] and then increase slowly in a broad region
of coupling strength [region II]. With the further increasing of
coupling strength, networks with fewer external links (larger
$Q_T$) will show degraded synchronization performance, while those
with more external links will be much better [region III]. After
that all the curves will reach steady states [region IV]. As for
the synchronization of individual community, $C_C$ will increase
monotonically with the increasing of coupling strength. However,
$C_C$ does not show a monotonic dependence on the number of
external links, as seen in Fig. \ref{Fig1}(b). We will investigate
the change of $C_C$ with respect to $Q_T$ in more detail later in
this paper. We also study the effects of external links on
dynamical modularity $Q_D$, which is obtained from the correlation
matrix $(c_{ij})$. Figure \ref{Fig1}(c) shows that dynamical
modularity $Q_D$ has two maxima for each curve as a function of
the coupling strength, one in region I at a very small coupling
strength and the other in region III.

\begin{figure}
\scalebox{0.45}[0.45]{\includegraphics{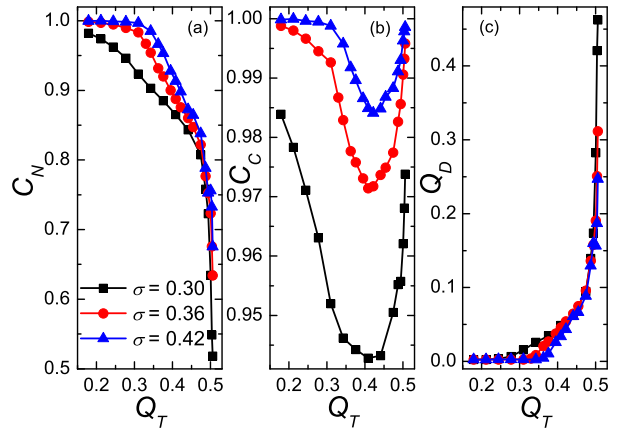}}
\caption{(Color on line) The change of $C_N$ (a), $C_C$ (b) and
$Q_D$ (c) as with respect to $Q_T$ at various coupling strength
$\sigma$ as indicated. The other network parameters are the same
as in Fig. \ref{Fig1}.}\label{Fig2}
\end{figure}

In the following we investigate into the mechanism of these
phenomena. As seen in Fig. \ref{Fig1}(d) for the region of small
couplings, $C_C$ grows faster than $C_N$, leading to enhanced
dynamical modularity. Then $C_N$ becomes large too and the
dynamical clusters become smeared and $Q_D$ is reduced. It worth
mentioning that very small coupling strength could make the
network go into very good phase synchronization state, and at the
coupling strength $\sigma=0.030$ the initial isolated subnetworks
will reach complete synchronization state. To show explicitly the
synchronization state of nodes in different communities, we
plotted the time series of the variable $x(t)$. In each community
two nodes (not directly connected) are selected, one having some
external links and the other one only having internal neighbors.
By comparing the four curves we found that phase synchronization
throughout the networks occurs at very small coupling strength
even for very few external links, as seen in Fig. \ref{Fig_1b}(a).
This corresponds to the slowly increasing region of $C_N$ [region
II in Fig. 1(a)]. With the further increase of the coupling
strength, the networks move to a regime of fierce competition: for
the networks with very few external links, individual community
get ready to reach complete synchronization state and the
synchronization of nodes in the same community is rather good, but
because of the random initial states different communities
oscillate at different phase, and the small number of external
links fail to bring the phases of the two communities to each
other, thus the global synchronization shows even worse
performance compared to weaker coupling strength. This phenomenon
can be seen in Fig. \ref{Fig_1b}(b). The good synchronization
performance within individual communities and weak synchronization
between them result in the prominent dynamical modularity (region
III, Fig. 1(a, c)). As for the networks with enough external
links, communities yield to the effects of external links, and
there is no significant difference between intra-community
synchronization and inter-community synchronization [Fig.
\ref{Fig_1b}(c)], leading to reduced dynamical modularity $Q_D$
(Fig. 1(c)). And when the number of external links is between
these two situations, the individual community tries to oscillate
independently but is inevitably affected by the others to some
extent, which results in degraded synchronization performance of
individual community. This is the reason for the decreasing of
$C_C$ at some topological modularity $Q_T$ (Fig. 1(b)). When the
coupling strength becomes stronger and stronger, the state
information exchanges more smoothly between different communities
and the states of different communities oscillate almost fully
synchronously (Figure \ref{Fig_1b}(c) ). The global
synchronization increases again and the dynamical communities
become not so prominent. This procedure of competition and
coordination can be further manifested by the state differences of
a node to its neighbors in the same community ($d_I$) and in the
other communities ($d_E$). Here the difference for a node $i$ and
its neighbors $j$ in a community $S$ is defined as
$d_i=\sqrt{\langle (\bar{x}_{j,S}-x_{i})^2\rangle}$, where
$\bar{x}_{j,S}$ is the average state of $i$'s neighbors in
community $S$ and $\langle \cdot \rangle $ denotes averaging over
time. $d_I$ and $d_E$ averaged over nodes are shown as functions
of connection strength $\sigma$ in Fig. \ref{Fig_1b}(d). It is
evident that $d_I$ almost keeps decreasing with $\sigma$, which
indicates that the state of a node gets close to its neighbors in
the same community gradually. However, the curve of $d_E$ has a
distinct peak at about $\sigma*=0.026$, which shows that in this
region of coupling strength the state of a node is far away from
its neighbor in the other community. In summary, we can see that
at very weak coupling strength, the external links could
effectively transmit the state information of different
communities to make the nodes of the whole network have the
similar phase, and at very strong coupling strength, the external
links could also be efficient to bring together the states of
different communities. However, when the coupling strength is
between these two cases, the synchronization in communities are
too strong individually and the external links fail to synchronize
them.

We also investigate $C_N$, $C_C$ and $Q_D$ as functions of the
topological modularity $Q_T$ with fixed coupling strengths (Fig.
\ref{Fig2}). Obviously, the synchronization performance of global
networks is degraded with the increase of $Q_T$, and dynamical
modularity $Q_D$ will keep about zero at small $Q_T$ and increase
sharply when the community structure becomes prominent enough at
large $Q_T$. Interesting behavior happens for synchronization of
individual community where $C_C$ display a pronounced minimum. The
change of $C_C$ can be understood as follows: when internal links
are rewired to external links in the beginning ($Q_T$ decreasing),
the number of internal links is reduced and the number of external
links is increased, synchronization within community becomes
weaker. In fact, only decreasing the number of internal links or
only increasing the number of external links will both reduce
$C_C$. At this stage, the individual communities still oscillate
largely independently ($C_C \sim 1> C_N$). When there are enough
external links, as part of the whole network the individual
communities are not prominent at all and it is no longer
independent and $C_C \approx C_N$, corresponding to $Q_D \approx
0$.

Therefore, the minimum of $C_C$ corresponds to a critical point
$Q_{TC}$. When $Q_T>Q_{TC}$, communities oscillate rather
independently and the communication between communities are weak,
and for $Q_T<Q_{TC}$, the communication between communities are
smooth but the dynamical independency of individual community is
not prominent. So the critical point represent a balance region
where the individual community is still clearly independent while
the information transmission remains effective between different
communities.

\begin{figure}
\scalebox{0.42}[0.42]{\includegraphics{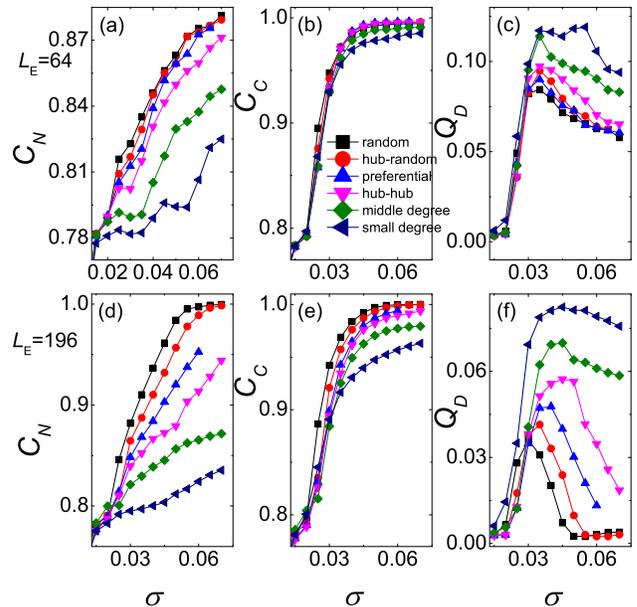}}
\caption{(Color on line) The change of $C_C$, $C_N$ and $Q_D$ with
$\sigma$ at $L_E=$64 and 196 for different rewiring strategies.
There are 2 communities in the network, each one is a BA network
with 100 nodes, and the average degree in each community is 16.
Each plot is obtained after the averaging of 50 network
configurations and 10 initial states of each
configuration.}\label{Fig_strategy}
\end{figure}

\section{Effects of connecting strategy of external links}

Intuitively, not only the number of external links but also the
connecting strategy of external links could affect the
synchronization performance of community networks. For example,
connecting nodes from different communities with larger degrees
seems to be more efficient for information exchange between the
communities, therefore, might be superior for global
synchronization. In this section we design several connecting
strategies to investigate whether it will be better for global
network synchronization to connect nodes with larger degree rather
than nodes with smaller degree.

To achieve our aim, we also take two BA scale-free networks as
communities and take the following connecting strategies: (i) keep
one end of external links on some nodes with largest degree (big
nodes) and connect the nodes with randomly selected nodes in other
communities (\emph{hub-random connection}), (ii) connect nodes
selected in different communities with probability $p\propto
k_i^4$, where $k_i$ is the degree of node belongs to the community
(\emph{preferential connection}), and (iii) connect big nodes
belong to different communities (\emph{hub-hub connection}).
Compared to the random connection, these three connecting
strategies make the distribution of external links biased more and
more to some big nodes. The selection probability in preferential
connection strategy guarantees that the big nodes of different
communities will not only connect densely among themselves, but
also connect to some \emph{smaller} nodes. Such a connection
scheme is consistent with the structure of brain cortex network
that the hub areas are almost fully connected among themselves to
form a hyper-community overlapping different communities
\cite{gorka}. Moreover, to compare the effects of degree of
connected nodes, we also designed the \emph{middle}/\emph{small
degree connection} strategy, which is characterized by connecting
nodes with middle/small degree in different communities. For all
the strategies, once the external link number is fixed, the
topological modularity is almost the same, but the synchronization
performances for networks connected with different strategies are
obviously different. Figure \ref{Fig_strategy} shows the
comparison results of these strategies. From the figure we can
conclude that the distribution of external links homogeneously on
nodes with different degree will lead to the best global
synchronization performance but the worst dynamical modularity. If
the external links must concentrate on several hubs, connecting
large nodes will be much better.

\section{Synchronization in modular cortical networks}

\begin{figure}
\epsfig{figure=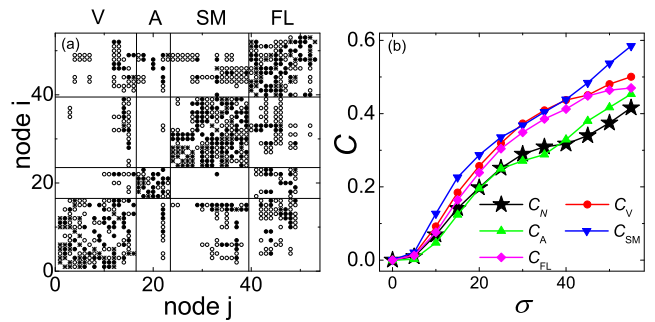, width=8.8cm} \caption{(Color online) (a) The
cortical network of cat. A node represents a functional region of
the cortex and a link represents the existence of fiber projection
between two regions. The different symbols represent different
connection weight: $1$ ($\circ$ sparse), $2$ ($\bullet$
intermediate) and $3$ ($\ast$ dense). The communities/functional
subdivision V, A, SM and FL of the network are indicated by the
solid lines. (b) The change of synchronization performance of
global network ($C_N$) and of the four communities ($C_V$, $C_A$,
$C_{SM}$ and $C_{FL}$) vs. the coupling strength $\sigma$. Each
plot is obtained after the averaging of 10 initial states.}
\label{Fig_cat}
\end{figure}

Modular organization is prominent and synchronization dynamics is
of special important for functioning in neural systems
\cite{claus,cs_1,cs_2,cs_NJP, bumull}.

The critical competition regime in the synchronization of
community network is of special importance from the viewpoint of
information processing, where the specialized processing using the
dynamical communities can co-exist with the global exchange of
information and integration from specialized communities. In
neural systems a combination of these two ingredients, the
functional segregation and integration, is believed to be crucial
to underly the structural and dynamical organization for effcient
and diverse functioning \cite{complexity,complexity1}. The
synchronization properties of the preferential connection scheme
as in the real cat cortical network allow quite strong global
synchronization and meanwhile maintain intermediate dynamical
modularity, again pointing to a meaningful balance between
segregation and integration.

In order to study the relevance of this interesting observation,
we study dynamics of realistic cortical neural network of cat. Cat
cortical network \cite{cat} [Fig. \ref{Fig_cat} (a)] has 53 nods,
826 weighted links and four communities that carry out the four
functions: visual (V), auditory (A), somato-motor (SM) and
frontolimbic (FL). In this network, each node represents a brain
area which are composed of huge number of interacting neurons. The
rhythmic activity of such neural ensemble can be modelled by
neurophysiologically realistic neural mass oscillators
\cite{neuronmass}. The dynamical equation of coupled neural mass
model is \cite{cs_NJP}
\begin{eqnarray}
  \ddot{v}^P_i&=&Aaf(v^E_I-v_i^I)-2a\dot{v}^P_i-a^2v^P_i, \nonumber\\
  \ddot{v}^I_i&=&BbC_4 f(C_3v^P_i)-2b\dot{v}^I_i-b^2v^I_i, \\
  \ddot{v}^E_i&=&Aa \Big [ C_2f(C_1v^p_I)+p_i(t)+\frac{\sigma}{\langle S\rangle}
  \sum^N_j W_{ij} \, f(v^E_j-v^I_j) \Big ] \nonumber\\
  &-&2a\dot{v}^E_j-a^2v^E_j, \nonumber
  \label{mass_eq}
\end{eqnarray}
where $v^P_i$, $v^I_i$ and $v^E_i$ are the average post-synaptic
membrane potentials of pyramidal cells, excitatory interneurons
and inhibitory interneurons of the area $i$, respectively. A
static nonlinear sigmoid function $f(v)=2e_0/(1+e^{r(v_0-v)})$
converts the average membrane potential $v$ into an average pulse
density of action potentials. Here $v_0$ is the post-synaptic
potential corresponding to a firing rate of $e_0$, and $r$ is the
steepness of the activation. The parameters $A$ and $B$ represent
the average synaptic gains, $1/a$ and $1/b$ the average
dendritic-membrane time constants. $C_1$ and $C_2$, $C_3$ and
$C_4$ are the average number of synaptic contacts, for the
excitatory and inhibitory synapses, respectively. $p_i$ represents
afferent inputs from subcortical systems. A more detailed
interpretation and the standard parameter values of this model can
be found in~\cite{neuronmass}. We take the parameters as in
\cite{neuronmass} so that the model generates alpha band periodic
oscillations. As in reference~\cite{neuronmass}, in our
simulations we take the subcortical input as $p_i(t)=p_0+
\xi_I(t)$ where $\xi_I(t)$ is a Gaussian white noise with standard
deviation $D=2$. $W_{ij}$ is the coupling strength from area $j$
to area $i$. We normalize coupling strength $\sigma$ by the mean
intensity $\langle S \rangle $ where the connection intensity
$S_i=\sum_j^N W_{ij}$ is the total input weight to node $i$.

In this system of noisy oscillations, we also use the correlation
between nodes to measure the network synchronization performance.
Fig. \ref{Fig_cat} (b) shows the change of $C_N$ and $C_C$ of the
four communities with the coupling strength $\sigma$. Unlike in
Fig. 1(a), here we do not observe the region III where $C_N$
decreases with $\sigma$; however, there is a regime at
intermediate couplings where $C_N$ increases slowly.

\begin{figure}
\scalebox{0.45}[0.45]{\includegraphics{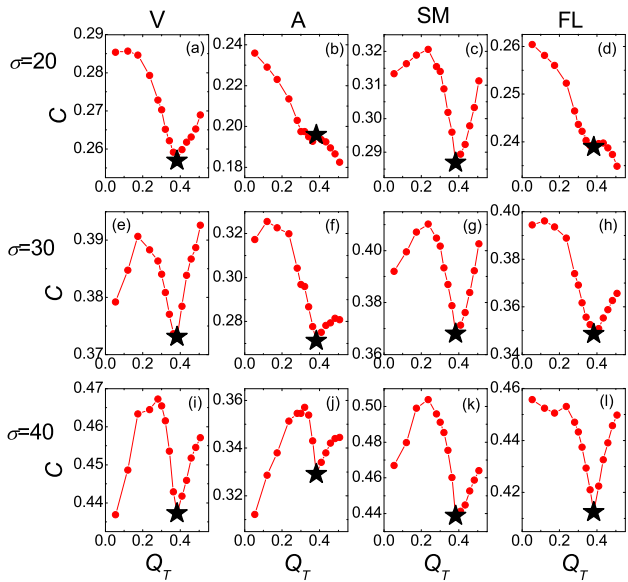}} \caption{(Color
on line) The change of $C_C$ vs. $Q_T$ at $\sigma=$20.0, 30.0 and
40.0. The stars present the results of the original cat cortical
network. Each other plot is obtained after the averaging over 50
network configurations and 10 initial states of each
configuration.} \label{CatC}
\end{figure}

In order to see if the real system could work in the
balance/critical states, we compare the synchronization
performance of the real network with rewired networks having
larger or smaller topological modularity. This is obtained by
rewiring the internal links to external links or rewiring the
external links to internal links of the original cat cortical
network. Figure \ref{CatC} shows the change of $C_C$ of the four
communities with respect to $Q_T$. Interestingly, in a broad
coupling strength region, the original network (star) is just at
the critical point of topological modularity where $C_C$ is a
minimum simultaneously in several or all the communities. These
results confirmed our expectation that the real cortical network
is such organized that it allows efficient global integration
under the condition that the functional segregation is also
maintained simultaneously. This finding provides a detailed
mechanism for our previous observation \cite{ZhaoM2010} that the
dynamical complexity measuring a balance between segregation and
integration is optimal in cat cortical networks.

\section{Discussion and Conclusion}

In summary, we have studied in detail the effects of external
links on the synchronization performance of community networks. We
have revealed an interesting competition between synchronization
of the global network and the individual communities. With the
increasing of the number of external links the global network
synchronization will be enhanced but the synchronization
performance of individual community will be degraded till a
critical point where the community structure is no longer
prominent. Afterwards, synchronization within the community
increases again as part of the global network. We also
investigated the impact of various connection strategies on the
global and community synchronization. We showed that connecting
nodes selected randomly in different communities will ensure
better global network synchronization, but weak dynamical
modularity. The distribution of the external links mainly among
the hubs nodes would allow both strong global synchronization and
clear dynamical modularity.

Interestingly, these discoveries in generic models are
demonstrated to be relevant in a realistic cat cortical network
with simulated neural population activities. Comparing the
synchronization properties to rewired networks with larger or
small modularity, we found that the real network is just at the
critical point. The hubs in this network are also responsible for
inter-community connections similar to the preferential connection
schemes in the model. Our analysis indicates that the cat cortical
network is such organized that it allows both segregated
performance with the communities and efficient integration of the
whole network.

Our work has provided a deeper understanding how the external link
number and connecting strategy affect the synchronization of
community networks as a whole as well as individual community.
These results not only present the possible reason for the real
cortical network to evolve to the current community structure from
the dynamical point of view but also provide useful methods to
regulate the synchronization of community networks for potential
applications.

\section{Acknowledgement}
This work is supported by the National Natural Science Foundation
of China (Grant No. 10805045), the key project of ministry of
education of China (Grant No. 210166), Guangxi Normal University
and also Hong Kong Baptist University and the Hong Kong Research
Grants Council (HKBU 202710).

\end{document}